\documentstyle[12pt,aasms4]{article}

\lefthead{Yasuhiro Shioya \& Kenji Bekki}
\righthead{Tightness of the color-magnitude relation}

\begin{document}
\title{Tightness of the color-magnitude relation of elliptical
galaxies and the epoch of major galaxy merging}

\author{Yasuhiro Shioya\altaffilmark{1} \& Kenji Bekki} 
\affil{Astronomical Institute, Tohoku University, Sendai, 980-8578, Japan} 

\altaffiltext{1}{Center for Interdisciplinary Research, Tohoku University, 
Sendai, 980-8578, Japan}

\begin{abstract}
We investigate a one-zone chemophotometric evolution model of 
disk-disk galaxy mergers in order to clarify whether or not 
galaxy mergers with the widely spread  merging epoch 
can reproduce reasonably well the observed small scatter of 
the color-magnitude ($C-M$) relation in  cluster ellipticals 
at low and intermediate  redshift ($z<1$). 
We consider that merger progenitor disks begin to consume 
interstellar gas at moderate rate from $z \sim 5$ and then 
merge to form an elliptical with the secondary starburst at 
$z=z_{\rm merge}$. 
We find that even if the epoch of galaxy merging is rather extended 
($0.3<z_{\rm merge}<3.0$), the dispersion in the rest-frame $U-V$ 
color among galaxy mergers is well within the observed one 
($\sim$ 0.05 mag at $z$ = 0).
We also find that the $z_{\rm merge}$ is required to be 
within a certain range to keep the observed  $C-M$ relation tight 
at a given $z$. 
For example, the required range of $z_{\rm merge}$ in galaxy mergers 
between Sa disks is $1.3<z_{\rm merge}<3.0$ for cluster ellipticals 
at $z$ = 0.895, $0.9<z_{\rm merge}<3.0$ for $z$ = 0.55, 
and $0.3<z_{\rm merge}<3.0$  for $z$ = 0. 
The main reason for the derived  small scatter is  that 
younger stellar populations, 
which are formed during the secondary starburst of galaxy mergers, 
are formed preferentially from more metal-enriched interstellar gas. 
This result reinforces the Worthey's suggestion (Worthey et al. 1996) 
that the age-metallicity conspiracy, 
which means that younger stellar populations are preferentially 
more metal-enriched, can operate to keep the tight $C-M$ relation. 
These numerical results imply that the observed small scatter 
in the $C-M$ relation at low and intermediate redshift ($z < 1$) 
{\it does not necessarily} require the coevality of 
elliptical galaxies in clusters or their formation at high $z$, 
which has been conventionally believed in the classical 
passive evolution picture. 

\end{abstract}


\section{Introduction}

Redshift evolution of fundamental physical relations in elliptical 
galaxies is generally considered to give strong constraints on 
the formation and evolution of elliptical galaxies. 
For example, evolution of the color-magnitude ($C-M$) relation 
with redshift ($z$) suggests that elliptical galaxies 
are old, coeval, and homogeneous systems passively evolving 
after the single initial burst of star formation associated 
with dissipative galaxy formation at $z > 2.0$ 
(Arag\'on-Salamanca et al. 1993; Ellis et al. 1997;
Stanford, Eisenhardt, \& Dickinson 1998).
This classical picture of coeval elliptical galaxy formation 
also appears to be supported by 
small redshift evolution  of both the mass-to-light-ratio 
(van Dokkum \& Franx 1996) and 
the $\rm Mg_{2} - \sigma$ relation (Ziegler \& Bender 1997). 
The considerably tight $C-M$  relation (Bower, Lucey, \& Ellis 1992) 
and the Fundamental Plane (e.g., Djorgovski \& Davis 1987) 
at the present  epoch,  
and redshift evolution of the slope and the zero-point of 
the $C-M$ relation 
(Kodama \& Arimoto 1997; Gladders et al. 1998; Kodama et al. 1998) 
furthermore seem to support 
the coevality of elliptical galaxy formation. 

An increasing  number of recent observational results, however, 
shed a strong doubt on this long-standing view of elliptical galaxy 
formation, and suggest that there is great variety of 
star formation history among elliptical galaxies, 
such as the epoch of major star formation, 
the duration and efficiency of star formation 
(Worthey, Faber, \& Gonzalez 1992; Faber et al. 1995; 
Worthey, Trager, \& Faber 1996).
In particular, Faber et al. (1995) suggested that the `apparent age 
spread', which is inferred from the combination of line index analysis of elliptical
galaxies, 
amounts to $\sim$ 10 Gyr.
Schweizer \& Seitzer (1992) found that in merger remnants with
morphologically fine structures, the last merging epoch, which corresponds
to elliptical galaxies formation, ranges from 4.6 Gyr to 8.0 Gyr ago.
These observed  spread in  `apparent mean age' seem to be inconsistent
with the aforementioned coevality of elliptical galaxy formation
expected mainly from the redshift evolution of the $C-M$ relation.

The purpose of this paper is to give a plausible answer to the above
apparent inconsistency in the epoch of elliptical galaxy formation.
We adopt the merger scenario of elliptical galaxy formation
(e.g., Toomre \& Toomre 1972)
and thereby investigate to what degree the difference in
the epoch of major galaxy merging (i.e., the epoch of elliptical
galaxy formation) can be allowed to preserve the observed 
small scatter of the $C-M$ relation of cluster ellipticals 
($\sim 0.05$ mag) at  $z$ = 0 (Bower et al. 1992),
 0.55 (Ellis et al. 1997), and 0.895  (Stanford et al. 1998). 
We find that owing to  the age-metallicity conspiracy proposed  by
Worthey et al. (1996), the observed small scatter in the 
$C-M$ relation  
can be reproduced reasonably well even in star-forming galaxy mergers
with the widely spread merging epoch. 
This result accordingly reinforces the recent results of 
Kauffmann \& Charlot (1998), 
in which the tight $C-M$ relation can be successfully
reproduced by merger scenario of elliptical galaxy formation
based on  a hierarchical clustering  scenario.
This result  furthermore implies   that the previously suggested interpretation
of the tightness of the $C-M$ relation at low and intermediate 
 redshift ($z < 1$)
is {\it not unique},  thus that 
the formation 
epoch of elliptical galaxies 
can be more
widely spread than the classical passive evolution picture predicts.
Thus, the above apparent inconsistency in the interpretation
of the $C-M$ relation can be due primarily  to the fact
that  previous studies claiming the coevality of elliptical galaxy formation
did not explore so extensively 
possible variety in  star formation history  of elliptical 
galaxies.

\section{Model}

 We adopt  a one-zone chemophotometric evolution model of 
 elliptical galaxies
 formed by major disk-disk galaxy mergers with the  
 merging  epoch widely spread, and thereby investigate
 to what degree the difference of the merging
 epoch can be allowed for keeping the observed tightness of the
 $C-M$ relation in cluster ellipticals at a given redshift.
 The remarkable  differences in model assumptions between the present merger model  and previous ones 
 based on a classical initial burst picture of elliptical galaxy formation (e.g.,
 Bower et al. 1992; Kodama \& Arimoto 1996) are
 the following two.
 The first is that the present model allows {\it continuous} and 
{\it moderate} star
 formation of galaxies before the onset of the secondary starburst
 associated with the elliptical galaxy formation via merger events.
 The second is that in the present model, the epoch of the formation of
 elliptical galaxies (more accurately, the formation
 of elliptical morphology) is assumed to be the epoch of the secondary starbursts
 in mergers. 
 These two differences generate qualitatively different results in the evolution
 of the $C-M$ relation between the present study and the previous ones
 based on a classical initial burst picture of elliptical galaxy formation. 
The elliptical galaxy formation with the secondary 
starburst has been investigated by Kauffmann (1996) 
in the context of the galaxy formation in the cold dark matter universe 
and by Charlot \& Silk (1994) in the context of the origin of 
Butcher-Oemler galaxies. 
We follow the chemical evolution of galaxies by using 
the model described in Matteucci \& Tornamb\`{e} (1987) 
which includes metal-enrichment processes of  
type Ia and II supernovae (SNIa and SNII). 
We adopt the Salpeter initial mass function (IMF), 
$\phi(m) \propto m^{-1.35}$, 
with upper mass limit $M_{\rm up}=120 M_{\odot}$ and 
lower mass limit $M_{\rm low}=0.1 M_{\odot}$ for most of the models.
We also investigate the models with the slope
of IMF, $x$, equal to 1.10 
in order to confirm that the results
derived for models with the Salpeter IMF can be generalized. 
The fraction of close binary stars (represented by  $A$ parameter
in Matteucci \& Tornamb\`{e} 1987),
which controls the frequency 
of SNIa relative to SNII, is assumed to be 0.1. 
To calculate the ejected mass of gas and heavy elements, 
we use stellar yields derived by  Woosley \& Weaver (1995) for SNII, 
Nomoto, Thielemann, \& Yokoi (1984) for  SNIa, 
and Bressan et al. (1993)  
and  Magris \& Bruzual (1993) for low and intermediate mass stars. 
We calculate photometric properties of galaxies as follows. 
The monochromatic flux of a galaxy with  age $T$,
$F_{\lambda}(T)$, is described as 
\begin{equation}
F_{\lambda}(T) = \int_0^T F_{{\rm SSP},\lambda}(Z,T-t) 
\psi(t) dt \; ,
\end{equation}
where $F_{{\rm SSP},\lambda}(Z,T-t)$ is  
a monochromatic flux of single stellar population (SSP) 
of age $T-t$ and metallicity $Z$, and $\psi(t)$ is
time-dependent star formation rate described later. 
In this paper, we use the spectral library 
GISSEL96 which is the  latest version 
of Bruzual \& Charlot (1993). 

 The star formation history of galaxy mergers is characterized by
 three epochs. 
The first  is $z_{\rm form}$ in redshift at which merger progenitor
 disk galaxies form and begin to consume 
 initial interstellar gas by star formation with the moderate
 rate and the value of $z_{\rm form}$ is fixed at 5.0 for all models.
 The second is $z_{\rm merge}$ at which two disks merge with each other
 and morphological transformation of galaxies (to ellipticals)
 and the accompanying starburst happen.
 The third is 
$z_{\rm end}$  at which star formation ceases and 
it is defined as the epoch at which stellar mass fraction becomes 0.9 
in our models. 
Star formation rates of galaxy mergers 
during $z_{\rm form} \le  z < z_{\rm merge}$,
 $z_{\rm merge} \le z<z_{\rm end}$,  and $z_{\rm end} \le z$ are described below.
 Throughout the evolution of galaxy mergers, the star formation rate is assumed
 to be proportional to gas mass fraction ($f_{g}$) of galaxies;
\begin{equation}
\psi(t) = k f_g \; , 
\end{equation}
 where $k$ is a parameter which controls the  star formation rate.
 This parameter $k$ is given as follows:
\begin{equation}
k = \left\{
\begin{array}{ll}
k_{\rm disk} \; \; \; & z_{\rm form} \le z < z_{\rm merge} \; ,\\
k_{\rm merge} \; \; \; & z_{\rm merge} \le  z < z_{\rm end} \; ,\\
0 \; \; \; & z_{\rm end} \le z \; .\\
\end{array}
\right.
\end{equation}
The parameter values of $k_{\rm disk}$ investigated in the present study
are 0.325 in units of Gyr$^{-1}$, 
which corresponds to an admittedly plausible star formation rate for Sa disks
(e.g., Arimoto, Yoshii, \& Takahara 1992),
0.225 (Sb), and 0.056 (Sc). 
These values of $k_{\rm disk}$ are consistent 
with the time scale of star formation estimated from 
observations of disk galaxies (e.g., Kennicutt et al. 1994). 
$k_{\rm merge}$ is fixed at 10.0, 
which is about two-order of magnitude larger than that typical values of $k_{\rm disk}$
adopted in the present study.
 The strength of the starburst  $k_{\rm merge}$ in the present model
 is consistent with
 observational results on starburst in gas-rich galaxy mergers 
(e.g., Sanders et al. 1988).

By using the above chemophotometric model, we investigate the evolution
of the rest-frame $U-V$ color in galaxy mergers with $z_{\rm merge}$
= 0.3, 0.5, 0.7, 0.9, 1.1, 1.3, 1.5, 1.6, 2.0 and 3.0 
for each Sa, Sb, and Sc model and 
$z_{\rm merge}=0.1$ for Sb and Sc model. 
The reason why we do not investigate the merger with 
$z_{\rm merge}=0.1$ for Sa model is that 
the fractional mass of stars in the Sa disk is larger than 0.9 
at $z=0.1$ (see the dotted line in Figure 1). 
For comparison, we also investigate chemophotometric evolution
of an isolated disk for Sa, Sb, and Sc models. 
In the followings, the cosmological parameters 
$H_0$ and $q_0$ are  set to be 65 km s$^{-1}$ Mpc$^{-1}$ and 0.05 
respectively, which means that the corresponding present age of
the universe is 13.8 Gyr.


\placefigure{fig-1}
\placefigure{fig-2}
\placefigure{fig-3}

\section{Results}

Figure 1 shows the time  evolution of the fractional mass of stars  
for 10 models  with  $0.3 \le z_{\rm merge} \le 3.0$ in the Sa model. 
As is shown in Figure 1, star formation proceeds with moderate rate 
before galaxy merging and then rapidly 
after the onset of the secondary starbursts of galaxy merging. 
This continuous star formation before elliptical galaxy formation 
(before galaxy merging) just characterizes 
the star formation history of the present model. 
Figure 2 describes the time  evolution of the rest-frame $U-V$ color 
for the 10 models 
(hereafter the $U-V$ color means the rest-frame $U-V$ color). 
From this figure, we can derive the following two qualitative results. 
Firstly, the $U-V$ color becomes red within $\sim$ 2 Gyr 
after the secondary starburst of galaxy mergers for each model. 
The $U-V$ color difference among these models is 
only $\sim$ 0.04 mag. 
Secondly, the epoch at which the $U-V$ color becomes red depends on 
$z_{\rm merge}$ in such a way that the epoch is earlier for models 
with larger $z_{\rm merge}$. 
These two qualitative results are found to be true 
for Sb and Sc models. 
Accordingly, it is clear that for a given 
star formation history of merger progenitor galaxies 
(for Sa, Sb, and Sc models), 
$z_{\rm merge}$ is required to be within a certain range for 
the observed tightness of the $C-M$ relation at a given redshift. 
We mention that 
the derived $U-V$ color in our merger models is  bluer than 
that of cluster ellipticals in Bower et al. (1992) 
owing to the smaller metallicity. 
The mass-weighted (luminosity-weighted) mean stellar metallicity 
of our models is about a half (one-third) of solar metallicity, 
although about a solar metallicity is needed to 
reproduce the observed color. 
This is essentially because we adopt 
the Salpeter IMF with $M_{\rm low}  = 0.1 M_{\odot}$ 
and the one-zone chemical evolution. 
If we adopt shallower slope of IMF (e.g., $x=1.10$) 
or larger values of $M_{\rm low}$ 
(e.g., $M_{\rm low} = 0.6 M_{\odot}$) 
and {\it nevertheless} use the GISSEL SSP 
(it is not reasonable
to use the GISSEL SSP, since we here do not assume the 
IMF adopted in GISSEL96), 
we can reproduce redder color typical for cluster ellipticals
in Bower et al. (1992). 
However, since main purpose of the present study is not 
to successfully reproduce the absolute magnitude of global colors
typical for cluster ellipticals 
but to explore the origin of the tight $C-M$ relation, 
it is not unreasonable to discuss the origin of the tight $C-M$ relation  
by using the merger models with rather bluer colors.
In the last paragraph of this section, we will show 
the $U-V$ color evolution of models with shallower IMF ($x=1.10$) 
and confirm that the tight $C-M$ relation is achieved 
independent of the adopted IMF.

Figure 3 shows the time evolution of the $U-V$ color for model 
with $z_{\rm merge}$ = 3.0, 1.6, 1.3, 0.9, 0.3 and 0.1. 
Here, we introduce the {\it minimum} $z_{\rm merge}$ 
required to keep the small scatter of the  $C-M$ relation 
at a given redshift ($z$=0.0, 0.55, and 0.895) 
for Sa, Sb, and Sc models. 
If the $z_{\rm merge}$ 
of galaxy mergers is larger than the {\it minimum} value, 
the color scatter of galaxy mergers at a given redshift can be 
smaller than the observed one. 
The {\it minimum} $z_{\rm merge}$ is, for example, 
1.3 (1.3, 1.6) at $z$=0.895, 
0.9 (0.9, 1.1) at $z$=0.55, 
and 0.3 (0.1, 0.3) at $z$=0.0, in Sa (Sb, Sc) models.
Figure 3 accordingly means,  for example, that for  Sa models with 
$1.3<z_{\rm merge}<3.0$, the dispersion  of the $U-V$ color is 
well within the observed one in cluster ellipticals 
at $z$ = 0.895 ($\sim$ 0.05 mag). 
This figure furthermore shows that for keeping 
the tightness of the $C-M$ relation at higher redshift, 
$z_{\rm merge}$ should be larger. 
The gas mass fraction that has already been converted into stars 
at $z \sim 3.0$ are 0.26 for Sa model. 
This result suggests that a substantial fraction of initial gas 
has not necessarily been converted into stars already in higher 
redshift, which is conventionally required in the passive 
evolution picture. 
As is shown in Figure 3, 
these results are  true for  Sb and Sc models, 
although the range of the required $z_{\rm merge}$ depends on 
star formation histories of Sa, Sb and Sc models in such a way that 
the required $z_{\rm merge}$ is appreciably larger for Sc models. 
For example, the {\it minimum} $z_{\rm merge}$ is 1.6 
at $z=0.895$ in Sc models although it is 1.3 in both Sa and Sb models. 
Thus, the scatter of the $U-V$ color in galaxy mergers 
with a wide spread in merging epoch is found to be rather small, 
which suggests that the tight $C-M$ relation at low and 
intermediate redshift does not necessarily require 
the coevality of elliptical galaxy 
or their formation  at high redshift.

\placefigure{fig-4}
\placefigure{fig-5}
\placefigure{fig-6}

In order to clarify the reason for  the  small 
$U-V$ color scatter derived 
in galaxy mergers with the widely spread merging epoch, 
we investigate  $\log {\rm [MgFe]} - \log {\rm H{\small \beta}}$ diagram 
for Sa models with $z_{\rm merge}$ = 3.0, 1.3, 0.9 and 0.3, 
for Sb models with $z_{\rm merge}$ = 3.0, 1.3, 0.9 and 0.1, 
and for Sc models with $z_{\rm merge}$ = 3.0, 1.6, 1.3 and 0.3. 
By using the $\log {\rm [MgFe]} - \log {\rm H{\small \beta}}$ relations 
of the GISSEL SSP with various age and metallicity, 
we can observe the stellar populations characterizing the photometric 
properties of galaxy mergers (e.g., Faber et al. 1995). 
Because of the smaller metallicity, 
the values of $[{\rm MgFe}]$ in our models is smaller than 
observed values (e.g., Faber et al. 1995; Kuntschner \& Davies 1997). 
We note again that this discrepancy might be removed 
by adopting the IMF with larger stellar yield 
or relaxing the one-zone chemical evolution. 
 As is shown in each panel of Figure 4, 
 the characteristic metallicity of the galaxy merger 
with younger characteristic age (with lower $z_{\rm merge}$),  
 is larger than that 
with older characteristic age (with higher $z_{\rm merge}$).    
 This is  because in the later starburst of galaxy mergers, 
 younger stellar populations are formed preferentially
 from  more metal-enriched interstellar gas.  
 Worthey et al. (1996) have already pointed out  that
 if younger stellar population are more metal-enriched (with the Worthey's
 law of $\Delta \log {\rm age} /  \Delta \log {\rm metallicity} = -3/2$),
the $C-M$ relation can be kept tight.
Accordingly,  
 the small scatter in the present merger
 model is closely associated with the 
 Worthey's `age-metallicity conspiracy'.
To clarify the effect of metallicity on the small scatter of 
the $U-V$ color, we calculate 
  the time evolution of the $U-V$ color for the  ``single metallicity'' models 
  with  $0.3 \le z_{\rm merge} \le 3.0$ in the Sa model. 
Here the ``single metallicity'' model means a model in which 
chemical evolution  is {\it not} solved but 
the star formation history 
is exactly the same as those  of a model  solving  chemical 
evolution fully. 
In this model, all stellar populations with different ages
in a galaxy have the same metallicity. 
In this calculation, we set $Z$ to be 0.006 
which is a typical value in our models. 
As is shown in Figure~5, the scatter of the $U-V$ color 
in the ``single metallicity'' model is appreciably
larger than that in the Sa models described before in Figure 2. 
We also note that the scatter becomes small more slowly 
in the ``single metallicity'' model than models including chemical evolution. 
This comparative experiment accordingly confirms
that `age-metallicity conspiracy'  plays a vital role in
keeping the $C-M$ relation tight (compare Figure 2 and Figure 5). 
This result furthermore provides a qualitative explanation 
for the reason why the present study allows the spread in the 
formation epoch of elliptical galaxies (more accurately, 
the epoch of the secondary starburst) 
whereas the previous observational ones 
(Bower et al. 1992; Arag\'on-Salamanca et al. 1993) do not. 
This is principally because in the previous studies, neither variety 
of star formation in elliptical galaxies 
nor the effects of chemical enrichment on spectroscopic 
evolution of galaxies are so fully investigated.
 Thus, it is demonstrated that 
 owing to the age-metallicity conspiracy, the scatter of the $C-M$ relation
 observed in cluster ellipticals at low and intermediate 
redshift  can be kept small, 
 even if elliptical galaxies are formed by disk-disk galaxy mergers 
 with the widely spread merging epoch.

Figure~4 furthermore shows an interesting behavior 
in redshift  evolution of line index  on 
$\log {\rm [MgFe]} - \log {\rm H{\small \beta}}$ diagram. 
As is shown in Figure~4 for Sa model, 
apparent age spread on $\log {\rm [MgFe]} - \log {\rm H{\small \beta}}$ diagram 
is smaller than the age spread of merging epoch 
(the epoch of elliptical galaxy formation) 
and becomes smaller as the age of burst populations becomes larger. 
For example, although the time spread between $z_{\rm merge}=3$ and 
$z_{\rm merge}=0.3$ is 7.6 Gyr, 
the apparent age spread in the diagnostic diagram is 
less than 2 Gyr at $z=0$. 
This result suggests that even if the epoch of major galaxy merging 
is rather spread, the age spread inferred from the  
$\log {\rm [MgFe]} - \log {\rm H{\small \beta}}$ diagram 
can be considerably smaller at lower $z$. 
This result thus implies that we can not necessarily 
confirm the coevality of elliptical galaxy formation 
even by using the $\log {\rm [MgFe]} - \log {\rm H{\small \beta}}$ 
diagram {\it at lower redshift}. 
The result in Figure~4 moreover provides an implication on the recent 
observational results concerning 
the $\log {\rm [MgFe]} - \log {\rm H{\small \beta}}$ diagram 
for cluster ellipticals at low redshift 
(Worthey et al. 1996; Kuntschner \& Davies 1997;  
Mehlert  et al. 1997). 
Kuntschner \& Davies (1997) found that in the Fornax cluster, 
age spread inferred from ${\rm [MgFe]} - {\rm H{\small \beta}}$ 
diagram and C4648 - H$\gamma_A$ diagram 
is considerably  small among elliptical galaxies.
Mehlert et al. (1997) also found that in the Coma cluster,
the age spread among massive elliptical galaxies is rather small. 
These two results are different from the results of 
Worthey et al. (1996) and Trager (1997) that 
both age and metallicity in elliptical galaxies 
can be rather spread in their samples,
which implies that the star formation history of elliptical galaxies
is different from clusters to clusters and from environments to environments.
Assuming that  elliptical galaxies are formed by galaxy mergers,
the result in Figure~4 can give the following explanation 
for the apparent difference in age distribution of galaxies 
between the above studies. 
The smaller  age scatter  inferred from 
$\log {\rm [MgFe]} - \log {\rm H{\small \beta}}$ diagram
for cluster of galaxies in Kuntschner \& Davies (1997) and 
Mehlert et al. (1997) reflects  the fact that {\it mean} epoch of the last
starburst associated with galaxy merging is relatively earlier whereas
the larger scatter in Worthey et al. (1996) reflects the fact that
the epoch of the last starburst in elliptical galaxies is  more widely spread owing to the larger spread
in the epoch of galaxy merging.
Future observational studies will assess the validity of this interpretation about
the diversity in the  properties of 
$\log {\rm [MgFe]} - \log {\rm H{\small \beta}}$ diagram
for elliptical  galaxies.

Lastly we present the results of models with the slope of 
IMF ($x$) equal to 1.10 in order to confirm 
that the above numerical results for models 
with smaller stellar yield 
can be applied to models with larger stellar yield 
which can reproduce the observed color of elliptical galaxies 
with larger metallicity. 
In this calculation, we also use the GISSEL96, 
although the IMF of these models is different from 
that adopted in GISSEL96. 
Figure~6 clearly demonstrates that even for models 
with larger stellar yield, 
the diversity in the epoch of galaxy merging does not 
introduce large scatter in the $U-V$ color, 
which implies that the aforementioned age-metallicity conspiracy
does not depend on the stellar yield 
(as a result, stellar metallicity) of galaxies. 
Assuming that there is a certain relation 
between the galactic luminosity and the mean stellar metallicity,
this result implies that tightness of global color in 
a {\it given} luminosity in the $C-M$ 
relation can be kept in elliptical galaxies formed by galaxy merging. 
This accordingly implies that even the slope of the $C-M$ relation 
can be kept after the number increase  of elliptical galaxies 
formed by galaxy merging in the $C-M$ relation. 
Merger progenitor disk galaxies are observationally revealed 
to have mass-metallicity relation 
(Zaritsky, Kennicutt, \& Huchra 1994) and color-magnitude relation
similar to that of elliptical galaxies (Peletier \& de Grijs 1997).
The present study together with these two  observational results
thus predict that even if a sizable fraction of elliptical galaxies are 
formed by galaxy mergers, the slope of the $C-M$ relation does not evolve so significantly 
with redshift.

\section{Discussion and conclusions}
 
The present study predicts that even if the epoch of major galaxy 
merging (i.e., the epoch of elliptical galaxy formation) is 
rather spread, both the tightness and the slope of the $C-M$ relation 
can be kept owing to the age-metallicity conspiracy originally proposed 
by Worthey et al (1996).
This result accordingly provides a heuristic explanation for 
the result of Kauffmann \& Charlot (1998) 
in which the tight $C-M$ relation has been already reproduced 
in the merger scenario of elliptical galaxy formation 
based on the hierarchical clustering model.
The conclusions derived in the present study however 
{\it seem} to be inconsistent with those derived in previous ones 
on the redshift evolution of the slope, zero-point and tightness of 
the $C-M$ relation of elliptical galaxies (Bower et al. 1992; 
Arag\'on-Salamanca et al. 1993; Kodama \& Arimoto 1997). 
In particular, the present numerical results {\it seem} to 
disagree with those of Kodama \& Arimoto (1997) 
and Kodama et al. (1998) (see also Gladders et al. 1998), 
which claim that the considerably less significant evolution 
of the slope of the $C-M$ relation rejects the age spread 
larger than 1 Gyr among elliptical galaxies. 
The apparent disagreement between the 
present study and the previous ones 
(e.g., Kodama \& Arimoto 1997; Kodama et al. 1998) 
is due essentially to the fact that the previous studies 
inevitably  have over-interpreted  the redshift evolution of the $C-M$ relation 
owing to the ad hoc assumption adopted in the previous studies. 
Although the previous studies are considerably sensible and valuable, 
it is important  to point out the ad hoc assumptions adopted in 
the previous studies and thereby clarify the reason why 
the present conclusions are not consistent with those derived by the 
previous studies of Kodama \& Arimoto (1997) and Kodama et al. (1998). 
The following three are the ad hoc assumptions which inevitably 
lead the previous studies  to draw the strong and general conclusion 
that formation of elliptical galaxies (especially in the cores of clusters) 
are {\it as a whole}  coeval 
and occurred at high redshift. 
First is that elliptical galaxies are formed by {\it only one} 
initial starburst.
Owing to this assumption, 
time evolution of global colors of elliptical galaxies 
depends exclusively on the epoch of initial burst of star formation 
(i.e., the epoch of elliptical galaxy formation 
in the previous  study). 
As a result of this, 
the age difference between elliptical galaxies 
(i.e., the difference of the epoch of elliptical galaxy formation 
in the previous studies)
can be more clearly reflected on the redshift evolution 
of the slope of the $C-M$ relation in the previous studies. 
Accordingly the observed less significant evolution of 
the slope of the $C-M$ relation is more likely to be interpreted 
as an evidence that supports the coevality of elliptical 
galaxy formation. 
It is certainly reasonable to claim that 
the observed evolution of the $C-M$ relation reject 
the `pure age' sequence model 
which demands that less luminous ellipticals 
have younger age. 
However, it seems not to be so reasonable to draw 
strong and general conclusion 
that elliptical galaxies are formed at $z>2$ 
{\it only} from the redshift evolution of the $C-M$ relation. 
Considering the first ad hoc assumption in the previous studies, 
what is more accurate and plausible interpretation 
on the observed evolution of the $C-M$ relation is just that 
the formation of {\it stellar populations} 
in {\it some} elliptical galaxies in the cores of 
{\it some} clusters (not the formation of galaxies 
with structural and morphological properties similar to 
those of ellipticals)
can be coeval and occurred at higher redshift ($z > 2$). 
The second is that an elliptical galaxy in a cluster of galaxies 
at higher redshift is a precursor of an elliptical galaxy 
in a cluster at lower redshift. 
The third is that a cluster of galaxies observed at higher redshift 
is a precursor of a cluster of galaxies at lower one. 
These two ad hoc assumptions actually enable us to discuss 
the origin of elliptical galaxies in a more general way and 
thus lead us to draw more general conclusions on the formation epoch 
of elliptical galaxies. 
However, since there are no observational evidences 
which can provide the firm physical basis 
for the above assumptions at least now, 
it is questionable to give any general conclusions on 
the coevality of elliptical galaxy formation. 
Thus, these three assumptions adopted in the previous studies 
inevitably lead them to provide the strong and general conclusion 
that formation of elliptical galaxies are 
coeval and occurred at higher redshift. 
 The present study, on the other hand, does not adopt the above 
 three ad hoc assumptions, and rather relaxes these assumptions. 
 Furthermore the present study instead allows both 
 continuous and moderate star formation   
 (not strong initial starburst) 
 and the secondary starburst associated with galaxy merging, 
 and assumes that the epoch 
 of morphological transformation (into ellipticals) 
 does not necessarily coincide with the epoch of galaxy formation 
 (i.e., the epoch when the star formation begins). 
 The evolution of the $C-M$ relation in the present study 
 consequently  does not depend so strongly on the difference 
 in the formation epoch between  elliptical galaxies 
 (i.e., the epoch of major 
 galaxy merging with the secondary starburst). 
 As a result of this, the present merger model predicts 
 that even if the formation epoch of elliptical galaxies 
 (i.e., the epoch of galaxy merging)
 are rather spread,
 both the slope and tightness of the $C-M$ relation can be kept. 
 Thus, the essential reason for the aforementioned  apparent disagreement
 on the coevality of elliptical galaxy formation
 is that the present study does not adopt the above three ad hoc assumptions
 whereas the previous studies do. 
 The interpretation on the redshift evolution of the $C-M$ relation in each model
 can depend strongly on the assumptions adopted by each model. 
 It is safe for us to say that it is not clear, at least now,
 which of the two different conclusions on the coevality of elliptical galaxy
 formation is more plausible and reasonable.
 However, considering the above three ad hoc assumptions adopted in the previous studies,
 what is more reasonable interpretation on the redshift evolution
 of the $C-M$ slope is that only {\it stellar populations} (not elliptical
 morphology) in {\it some} ellipticals
 located in  the cores of {\it some} clusters of galaxies are formed at higher redshift. 
 We should not draw any {\it general} conclusions from the redshift evolution
 of the slope of the $C-M$ relation.

   Environmental difference of stellar populations (in particular,
   the existence of intermediate-age population) 
   in early-type galaxies has been already indicated by a number
   of observational studies (e.g., Bower et al. 1990; Rose et al. 1994;
   Mobasher \& James 1996).
   On the other hand, the tightness and the slope of the $C-M$ relation
   of early-type galaxies
   are observationally revealed not to depend so strongly on
   galaxy environments.
   These two apparently inconsistent observational results 
   on spectrophotometric properties of elliptical galaxies
   have called into the following question:
   ``Why does not   the $C-M$ relation of early-type galaxies
   depend strongly on galaxy environments (e.g., between
   rich clusters and poor ones), though stellar populations
   and star formation histories in early-type galaxies
   probably depend on 
   galaxy environments?''
   To give a plausible answer for this question seems to 
be important 
   because the above apparently inconsistent observational
   results give us valuable information both on the environmental
   difference in the details of physical processes of elliptical galaxy
   formation and on a certain mechanism for the tight $C-M$ relation. 
   However, no extensive theoretical studies have yet addressed the above 
   important question.
   The present study has shown that the age-metallicity conspiracy,
   which is achieved by younger and more metal-enriched stellar populations
   created in the secondary starburst of galaxy mergers,
   allows both the apparent age spread of elliptical galaxies
   and the tightness of the  $C-M$ relation.
   This result seems to  provide a clue to the above question. 
Since the {\it real} question concerning the tight $C-M$ relation
   is not to determine the typical
   epoch of elliptical galaxy formation but to give a convincing
   explanation for the reason why possible diversity in star formation
   histories of elliptical galaxies can allow the tight $C-M$ relation,
   more extensive theoretical studies including more variety of
   star formation history of elliptical galaxies and its likely
   dependence on galaxy environments are certainly worth for our
   deeper understanding of the origin of the tight $C-M$ relation.

 The present numerical results  are consistent with recent observational
 results which suggest that coeval elliptical galaxy formation with initial
 starburst at higher redshift ($z >  2.0$) is not promising.
 Kauffmann, Charlot, \& White (1996) revealed that only about one-third
 of bright E/S0 galaxies in the sample of Canada-France Redshift Survey
 were already in the passive evolution phase at $z \sim 1.0$.
 Franceschini et al. (1997)
 found a remarkable absence of early-type galaxies
 at $z > 1.3$ 
 in the $K$-band selected sample of early-type galaxies in
 the Hubble Deep Field (HDF), 
 which suggests either that early-type galaxies are formed by galaxy 
 merging with less prominent star formation or that 
 a dust-polluted interstellar gas obscures forming elliptical
 galaxies till $z = 1.3$.
 Zepf (1997) demonstrated that strong deficit of galaxies with
 extremely red colors in the HDF 
 means that the formation epoch of typical elliptical galaxies 
is   $z < 5.0$.
Sample galaxies in these studies are selected from {\it field ellipticals},
which possibly have star formation histories different from those of {\it cluster ellipticals}. 
Accordingly, it might not be plausible to derive strong conclusions on the formation
epoch of ellipticals. 
 However these observational results together with the present 
results seem to support the merger scenario
 which can naturally predict that the epoch of elliptical galaxy formation
 is rather extended ranging from high redshift to moderate one.

 Thus we  have succeeded in pointing out  that even if the epoch of
 elliptical galaxy formation 
(i.e., the epoch of major disk-disk galaxy merging, in this study)
 is rather widely spread, the tightness of the $C-M$ relation at low 
 and intermediate 
 redshift can be kept
 reasonably well.
 This result suggests  that coevality of elliptical galaxy formation,
 which has been  conventionally believed in the classical passive evolution picture,
is {\it not unique} interpretation  for the small  scatter of
 the $C-M$ relation. 
 This  furthermore implies that 
 {\it only} the tightness of the $C-M$ relation at a given redshift
 {\it does not necessarily} give strong constraints on the formation
 epoch of elliptical galaxies.
 Worthey et al. (1996) have already pointed out that the age-metallicity conspiracy
 can keep both the tightness of  the Fundamental Plane and that of
 the $C-M$ relation
 in elliptical galaxies.
  The present numerical study, which is different from the
  Worthey's single stellar population analysis,
  has confirmed that the proposed age-metallicity conspiracy can actually
  operate to keep 
convincingly 
the tightness of the $C-M$ relation
  of ellipticals formed by  disk-disk galaxy mergers.
  The present chemophotometric evolution model is, however,
  not so elaborated and realistic in that this model
  neither includes continuous gas accretion/merging
  expected from a specific
  cosmology (e.g., Baugh, Cole, \& Frenk 1996; Kauffmann \& 
  Charlot 1998) nor considers important dynamical
  effects of galaxy merging on chemical and photometric evolution
  of galaxies (Bekki \& Shioya 1998).
  Accordingly  it is our future study to confirm that 
  the results derived in the present preliminary study 
  can hold even for more sophisticated and realistic merger models.
 Furthermore, 
 we should  check whether or not observed redshift
 evolution of other fundamental relations such as the 
$\rm  Mg_{2} - \sigma$
 relation (Ziegler \& Bender 1997),
  the Fundamental Plane (van Dokkum \& Franx 1996),
  and the abundance ratio of [Mg/Fe]
 can be also reproduced self-consistently by our future  
 merger model.

\acknowledgments

K.B. thanks to the Japan Society for Promotion of Science (JSPS) 
Research Fellowships for Young Scientist.

\clearpage

\figcaption{Time evolution of the fractional mass of stars  for Sa models with different
$0.3 \le z_{\rm merge} \le 3.0$.  For comparison, the results of an
isolated Sa model is given by a dotted line.  Solid lines,
short-dashed lines, and long-dashed lines represent Sa models with
$1.0 < z_{\rm merge} \le 3.0$, models with $0.5 < z_{\rm merge} \le
1.0$, and models with $0.0 < z_{\rm merge} \le 0.5$, respectively.
\label{fig-1}}

\figcaption{Time evolution of the $U-V$ color for Sa models with different
$0.3 \le z_{\rm merge} \le 3.0$.  For comparison, the results of an
isolated Sa model is given by a dotted line.  Solid lines,
short-dashed lines, and long-dashed lines represent Sa models with
$1.0 < z_{\rm merge} \le 3.0$, models with $0.5 < z_{\rm merge} \le
1.0$, and models with $0.0 < z_{\rm merge} \le 0.5$, respectively.
The transient decrease of the color represents the epoch of secondary
starbursts for each merger model.  Note that the $U-V$ color becomes
red at the earlier epoch of galaxy evolution for models with larger
$z_{\rm merge}$.  We should emphasize here that because of the adopted
Salpeter IMF with $M_{\rm low} = 0.1 M_{\odot}$ (which is the only IMF
available for photometric and spectroscopic calculation in the GISSEL
SSP), the derived color is rather blue compared with that of cluster
ellipticals in Bower et al. (1992).  
\label{fig-2}}

\figcaption{Time evolution of $U-V$ color of mergers with $z_{\rm
merge}$ = 3.0 (solid), 1.6 (dotted), 1.3 (short-dashed), 
0.9 (long-dashed), 0.3 (dash-dotted) and 0.1 (dashed chain) 
for Sa models (upper panel), for Sb models (middle), 
and for Sc models (lower).  
For comparison, the result for an
isolated disk model is also given by a dotted line in each panel.  The
observed scatter of the $U-V$ color ($\sim$ 0.1 mag), which
corresponds to two times the 1 $\sigma$ dispersion ($\sim$ 0.05 mag),
is given by error bars for $z$ = 0.0, 0.55, and 0.895.  This figure
implies, for example, that for Sa models, the $z_{\rm merge}$ is
required to be $1.3 \le z_{\rm merge} \le 3.0$ for keeping the
observed tightness of cluster ellipticals at $z$ = 0.895, $0.9 \le
z_{\rm merge} \le 3.0$ for $z$ = 0.55, and $0.3 \le z_{\rm merge} \le
3.0$ for $z$ = 0.0.  Note that for keeping the small scatter at larger
$z$, the $z_{\rm merge}$ should be larger.  Note also that for Sc
models, the range of the $z_{\rm merge}$ for the tightness of the
$C-M$ relation at each redshift is relatively smaller.  \label{fig-3}}

\figcaption{The $\log {\rm H}{\beta}$-$\log {\rm [MgFe]}$ diagram for
Sa models with $z_{\rm merge}$ = 3.0, 1.3, 0.9, and 0.3 (upper), for
Sb models with $z_{\rm merge}$ = 3.0, 1.3, 0.9, and 0.1 (middle), and
for Sc models with $z_{\rm merge}$ = 3.0, 1.6, 1.3, and 0.3 (lower).
In each panel, vales of $\log {\rm H}{\beta}$ and $\log {\rm [MgFe]}$
at $z$ = 0.00, 0.55, and 0.895 are given by filled circles for models
keeping the tight $C-M$ relation at $z=0.895$ (left), 0.55 (center)
and 0.0 (right).  For clarity, the results at the same redshift are
connected by long-dashed lines.  For comparison, the results expected
from the SSP of the GISSEL96 are also given by dotted lines with open
circles in the same diagram for different age (3, 7, 11, and 15 Gyr)
and metallicity ($Z$ = 0.004, 0.008, 0.02).  Numbers plotted near open
circles represent the age of the SSP in units of Gyr.  This diagram
can represent the luminosity-weighted mean age and metallicity of
stellar populations in galaxies.  Note that galaxy merger dominated by
younger stellar populations are more metal-enriched.  This result
clearly demonstrates that owing to the Worthey's age-metallicity conspiracy,
the $C-M$ relation can be kept tight even in galaxy mergers with the
widely different merging epoch.  \label{fig-4}}

\figcaption{The same as Figure 2 but for the ``single metallicity'' Sa  model.  
Here the single metallicity model means a model in which 
chemical evolution  is {\it not} solved but the star formation history 
is exactly the same as those  of a Sa model  which solves chemical
evolution and is described in Figure 2. 
In this single metallicity model, all stellar populations with different ages
in a galaxy have the same metallicity.
For comparison, the results of an
isolated single metallicity Sa model is given by a dotted line.  Solid lines,
short-dashed lines, and long-dashed lines represent single metallicity 
Sa models with $1.0 < z_{\rm merge} \le 3.0$, 
models with $0.5 < z_{\rm merge} \le 1.0$, 
and models with $0.0 < z_{\rm merge} \le 0.5$, respectively.
\label{fig-5}}

\figcaption{The same as Figure 2 but for Sa  model
with the slope of IMF equal to 1.10.
For comparison, the results of an
isolated Sa model is given by a dotted line.  Solid lines,
short-dashed lines, and long-dashed lines represent Sa models with
$1.0 < z_{\rm merge} \le 3.0$, models with $0.5 < z_{\rm merge} \le
1.0$, and models with $0.0 < z_{\rm merge} \le 0.5$, respectively.
Note that compared with the results of models with the Salpeter IMF
in Figure 2,
the color for each model is appreciably redder. 
\label{fig-6}}

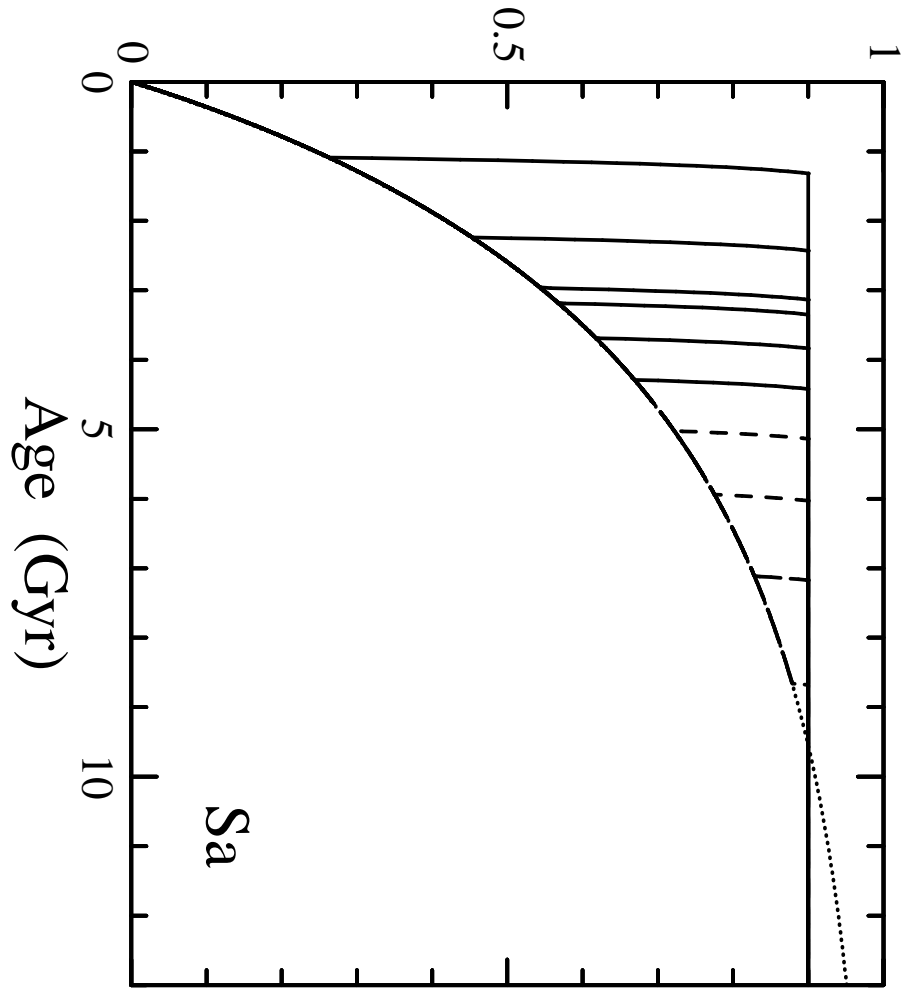
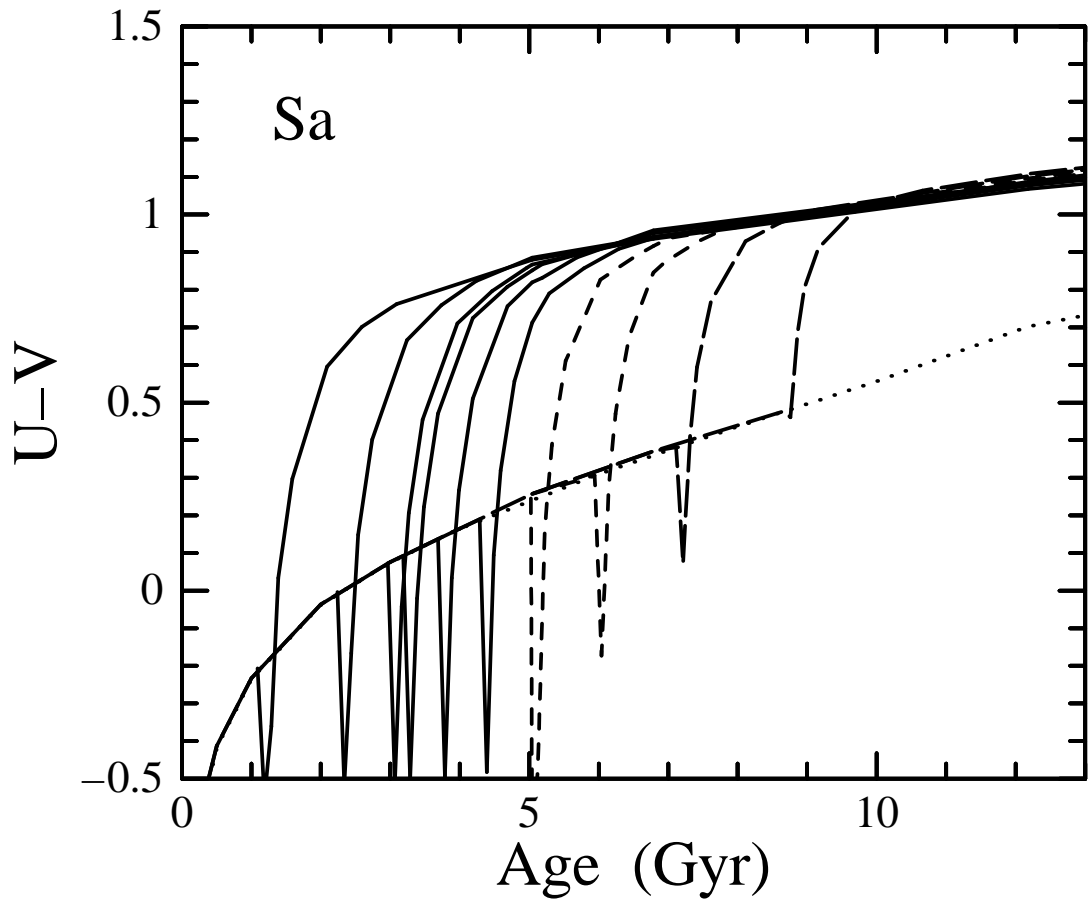
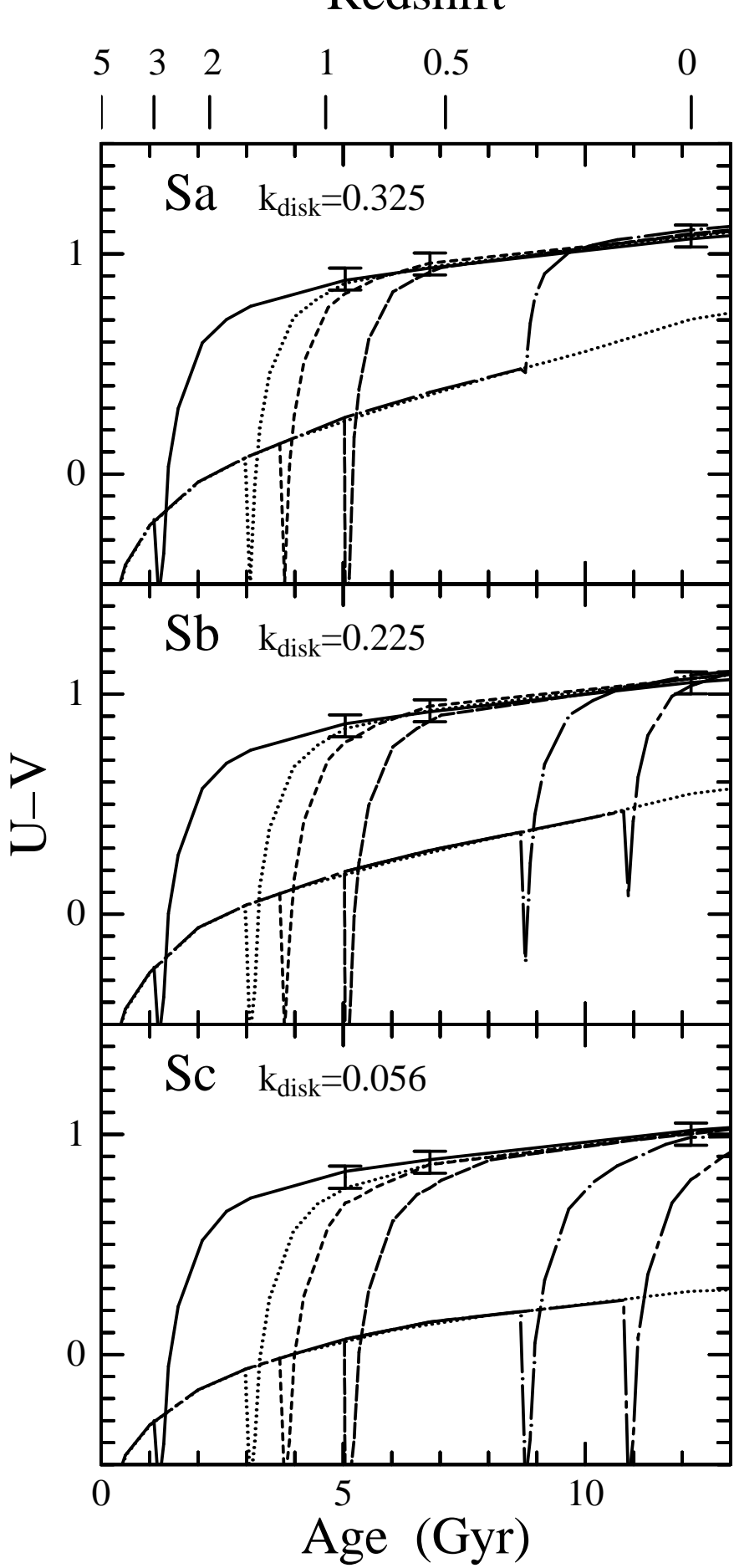
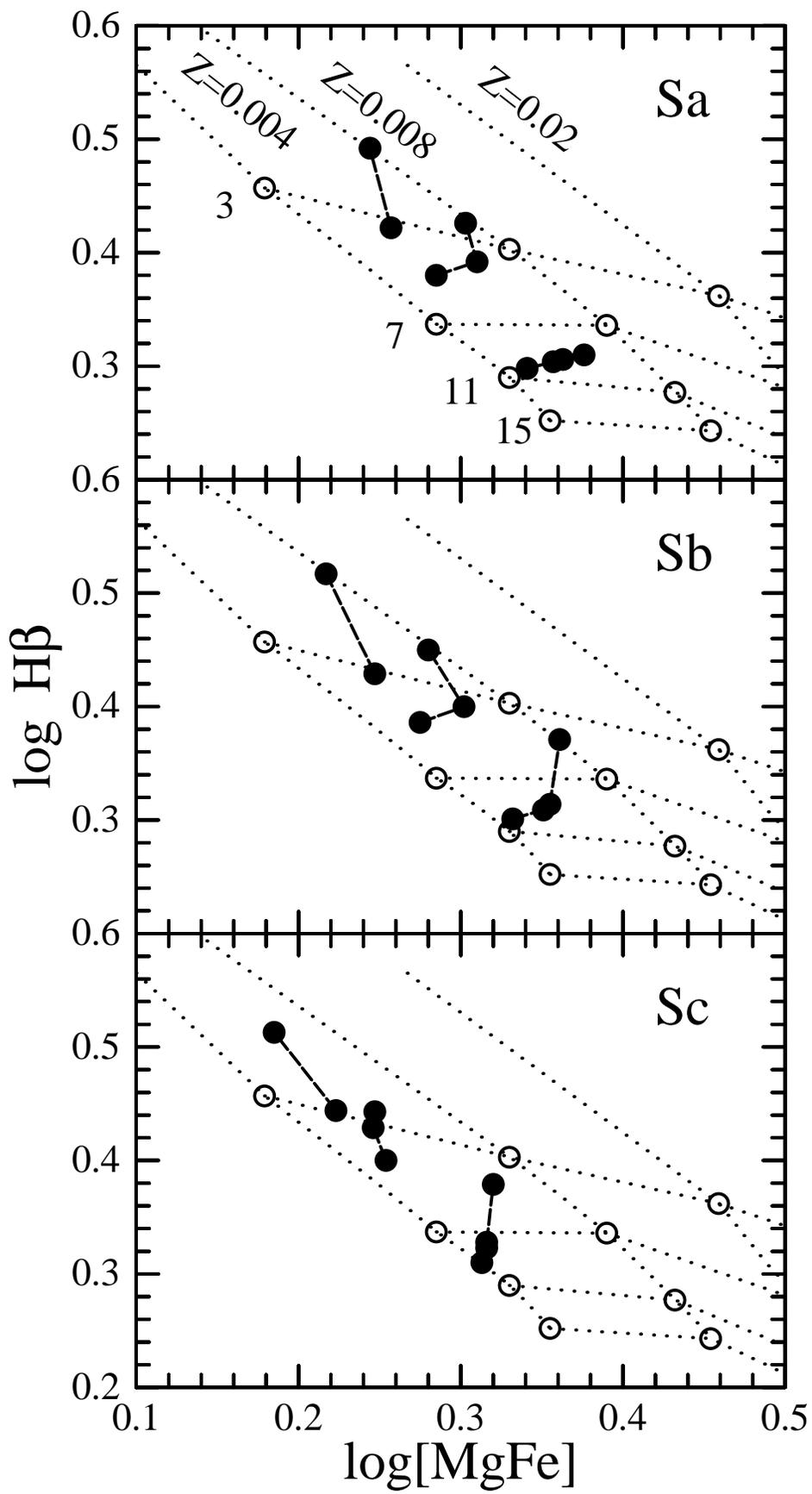
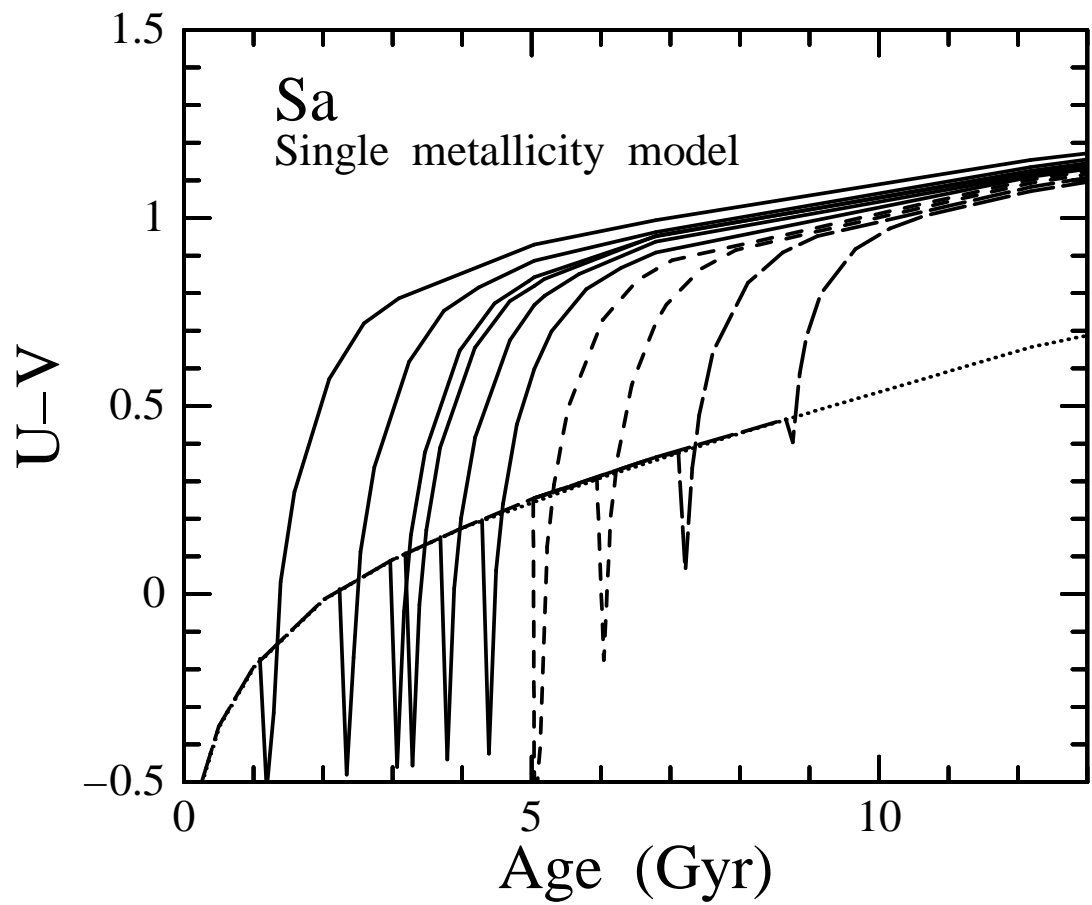
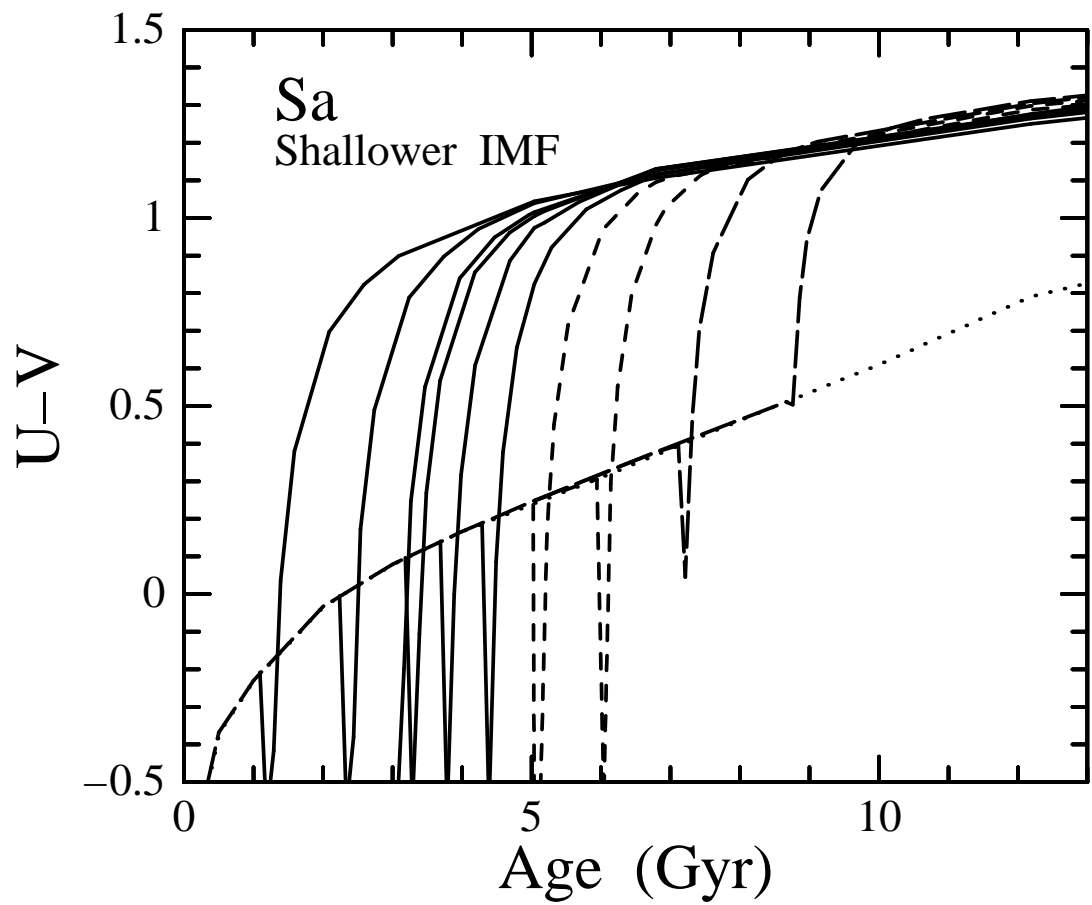
\end{document}